\documentclass[a4paper,11pt]{article}
\usepackage{times}
\usepackage{epsfig}
\usepackage{multirow}
\usepackage{fontenc}

\begin{document}

\title{\bf Social Media and Information Overload: Survey Results}

\author{Kalina Bontcheva, Genevieve Gorrell, Bridgette Wessels}

\date{}

\maketitle

\begin{abstract}

A UK-based online questionnaire investigating aspects of usage of
user-generated media (UGM), such as Facebook, LinkedIn and Twitter,
attracted 587 participants. Results show a high degree of engagement
with social networking media such as Facebook, and a significant
engagement with other media such as professional media, microblogs and
blogs. Participants who experience information overload are those who
engage less frequently with the media, rather than those who have
fewer posts to read. Professional users show different behaviours to
social users. Microbloggers complain of information overload to the
greatest extent. Two thirds of Twitter-users have felt that they
receive too many posts, and over half of Twitter-users have felt the
need for a tool to filter out the irrelevant posts. Generally
speaking, participants express satisfaction with the media, though a
significant minority express a range of concerns including information
overload and privacy.

Keyword list: User-Generated Media, Social Networking, Information Overload, Information Overlook, Facebook, LinkedIn, Twitter
\end{abstract}




\section{Introduction}\label{sec:intro}

User-Generated Media (UGM) are web services in which users share
information and provide web content of various forms for the benefit
of other users. Examples include Facebook, LinkedIn and Twitter. Due
to the recent explosion of interest in these media, more people than
ever before are publishing content on the internet. This makes a very
large amount of information available, and makes the problem of
consuming this information intelligently a pressing one for many
people.

UGM generally feature some means of connecting to other users, either
to document the relationship or to indicate an interest in what they
have to say. Users may then create posts or status updates, and the
service will provide to users the posts of those users to whom they
are connected. Variations from this formula exist, but for the most
popular services, this encapsulates the principle. Different UGM can
be characterised in different ways, describing their aims and focus as
positions on a spectrum. Interest-graph media~\cite{Ravikant2010}
encourage users to form connections with others based on shared
interests, regardless of whether they personally know the other
person. They aim to provide the information to users that they will
find most interesting. Twitter, a microblogging service in which users
share short status updates, encourages this model, and the "following"
relationship is often one-way, as a well known or particularly
interesting person will attract large numbers of followers who are
interested in what they have to say without themselves returning the
compliment in most cases. Social-graph media encourage users to
connect only with people they have real-life relationships
with. Facebook provides a way for people to keep in touch with friends
they might otherwise not talk to much, or share information between
friends who see a lot of each other. LinkedIn aims to provide an
introductions service in the context of work, where connecting to a
person implies that you vouch for that person to a certain extent, and
would recommend them as a work contact for others. Status update media
encourage short contributions from users, often outlining current
events in their lives or linking to something on the internet that
they think others might enjoy. These status updates are combined into
a time-ordered stream for each user to read. Blogs usually invite
longer contributions. Readers might comment on these contributions,
and some blog sites create a time stream of blog articles for
followers to read.

\begin{figure}[]
	\centering
	\caption{Number of Facebook Users by Year}
	\includegraphics{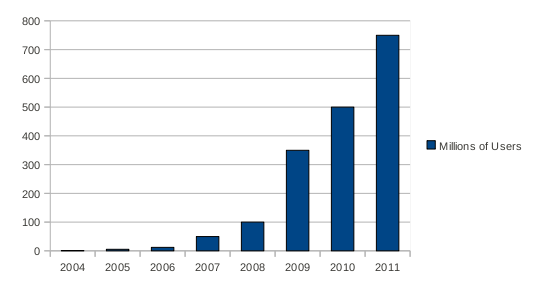}
	\label{fig-facebook-growth}
\end{figure}

As more users are drawn to a particular system, the growth rate seems
to increase. Consider the growth of Facebook as the most striking
example. As Figure~\ref{fig-facebook-growth} shows, prior to 2007,
uptake was modest. From 2007 onwards, however, an explosion in the
number of users took place, until now, in late 2011, the user base is
more than twice that of the United States of America, the third most
populous country in the world. Twitter shows a similar growth
pattern~\cite{Cheng2009}.

There are reasons why a snowball effect might be taking
place~\cite{Pankaj1990335}. As more people congregate on a particular
system, others may feel more incentive, or pressure, to join
too. There is more utility to using the system on which one can reach
the greatest number of relevant people. People may put pressure others
to join the system they use in order to increase their own utility on
that system. This leads to a tendency for one player to dominate
within each field, though applications making it very easy to partake
in multiple media may counter this tendency as it becomes easy for a
person to join several networks and have their posts automatically
sent to all their networks.

These media have the effect of making the world seem a much smaller
place, in terms of ease of communication with others regardless of
geographical location. With this abundance of access comes a greater
need to process the information input to make sense of
it. "Information overload" describes the difficulty a person can have
understanding an issue and making decisions in the presence of too
much information. The term was popularised by Alvin Toffler in his
bestseller, "Future Shock"~\cite{Toffler1984}. Part of the problem may
come from a diminishing of the signal-to-noise ratio, a pollution of
information with misleading, irrelevant or low quality
material. Another aspect comes from human cognition and approaches to
decision-making. We might assume that more information is always
better, and these media provide large quantities of time-stamped
material we feel compelled to keep up with for fear of missing
anything. We may have difficulty prioritising based on our values and
interests in the presence of so much information, always distracted by
that new piece of information that due to recency of exposure, seems
so important~\cite{jacoby1974}. A broader definition of information
overload, however, might simply be that a person finds it difficult to
keep up with the information they have chosen for themselves and often
fails to find time to attend to the quantity of information that they
feel they ought. Used in this way, the problem conceptualisation is
not new; as early as 1985, researchers began to postulate a problem
with computer-mediated communication systems such as email and message
board systems, which so entirely changed communication compared with
what had gone before~\cite{Hiltz:1985:SCC:3894.3895}. It is this sense
of the term that we focus on here, extended to a yet more recent
communication paradigm.


\begin{quote}There is no such thing as information overload, there's only filter failure\end{quote}

As this quote by Clay Shirky suggests, the profusion of information in
itself isn't the problem. More information, technically speaking,
ought to be a good thing, but to utilise it effectively, and protect
ourselves from its hazards, we need to filter better. To take
advantage of the unprecedented information availablity, we need
automatic processing of some kind. In this sense, information overload
is more of an opportunity than a problem. We have an abundance of
informational riches, but we need to develop filtering strategies in
order to turn it to our advantage.

The research we describe here provided an opportunity to see how usage
relates to different aspects of the media; for example, blogs and
microblogs are united by their focus on the interest graph. What
similarities will we see in the way people respond to these media? SNS
and microblogs focus on short status updates. In what ways will this
similarity unite usage patterns? The main focus of our research
however is on the information overload problem. To what extent are
people aware of this as a problem? What do they think should be done
about it? Who experiences information overload? Who doesn't? Does it
attach more to some media than others, and what are the
characteristics of those media?


\section{Related Work}\label{sec:related-work}

Literature regarding UGM is necessarily very recent, since the picture
is evolving so rapidly and findings true in one year may not be the
year after. Twitter, the microblogging service, has existed since
2006, and a rapid escalation of popularity has led to its having 300
million users just five years later. Its history indicates something
of how its current usage deviates from the use intended when it was
created. The first prototype was rolled out as an internal company
service~\cite{Arrington2006}. Later, expansion was fuelled as it came
into its own as a way to facilitate communication between attendees at
the SXSW festival~\cite{Douglas2007}. Tweeters used the service to
communicate about current events personally relevant to those
reading. The earlier Twitter tagline of "What are you doing" invites
primarily me-focused tweets; later this became "what's happening"
perhaps to reflect a broadening usage~\cite{Ehrlich2010}. Ehrlich and
Shami~\cite{Ehrlich2010} explore how company internal microblog usage
(BlueTwit) differs from current Twitter usage across the same group of
users, and find that company internal microblog usage focuses more on
communication such as question asking, directed posts and directed
questions and less on providing information and status
updates. However, "what's happening" may still not reflect how Twitter
is being used. Ehrlich and Shami found over 25\% of tweets in their
sample were directed posts, making Twitter into a short, public
message service, in addition to a means of sharing status updates
(11\%) and information (29\%).

Zhao and Rosson~\cite{Zhao:2009:WPT:1531674.1531710} emphasise the
"water cooler conversation" angle, again suggesting people tweet
within cliques. However in 2010, Kwak \textit{et
  al}~\cite{Kwak:2010:TSN:1772690.1772751}, in a much larger study of
106 million tweets, find metrics suggesting a widespread deviation
from social network-style usage. In particular, reciprosity is low,
suggesting that people are "following" others based on the often
one-way "interest graph"~\cite{Ravikant2010}, that is to say, linking
based on shared interests, rather than the by-definition two-way
relationship of social acquaintance ("social graph"). Trending topics
tend to be headline news or persistent news. "Retweeting", that is,
repeating a tweet with acknowledgement, is responsible for rapid
spreading of headlining information. Thus Kwak \textit{et al} suggest,
essentially, that Twitter is now a news medium. Particularly
interesting is the emerging possibility of crowdsourcing news via
microblogging services, capitalising on the fact that a major use of
microblogging is to talk about significant events, and such news will
tend to be passed on. Sakaki \textit{et
  al}~\cite{Sakaki:2010:EST:1772690.1772777} explore this possibility
in the context of earthquake and typhoon detection.

Social-graph services such as Facebook and LinkedIn enforce
relationship reciprocity by requiring that both parties consent to
being linked. By \textit{not} doing this, Twitter has allowed a
fascinating evolution to take place, from a social-based small
messaging service to a highly versatile social environment, finding
its niche as the main interest-graph medium but not limited to
that. Naaman \textit{et al}~\cite{Naaman:2010:RMM:1718918.1718953}
found over 40\% of their sample of tweets were "me now" messages, that
is, posts by a user describing what they are currently doing, in the
manner encouraged by the "what's happening" prompt. Next most common
were statements and random thoughts, opinions and complaints and
information sharing such as links, each taking over 20\% of the
total. Less common tweet themes were self-promotion, questions to
followers, presence maintenance e.g. "I'm back!", anecdotes about
oneself and anecdotes about another. Messages posted from mobile
devices are more likely to be "me now" messages (51\%). Females post
more "me now" messages than males. A relatively small number of people
undertake information sharing as a major activity; users can be
grouped into "informers" and "meformers", where meformers mostly share
information about themselves. Informers and meformers differ in
various ways. Informers tend to be more conversational and have more
contacts.

As well as enforced reciprocity, Facebook and LinkedIn differ from
Twitter in that they involve a privacy system. Whereas all tweets are
public, posts to Facebook and LinkedIn can be seen only by those to
whom permission has been granted. Despite this, privacy is more of an
issue among Facebook users than Twitter users, perhaps because some
kind of decision-making process is entailed. Hoadly \textit{et
  al}~\cite{Hoadley201050} investigate the reaction among users when
Facebook, in 2006, introduced the news feed, in which information
users previously would have had to seek out by going to each others'
pages was now aggregated into a time-ordered digest and placed front
and centre on the site. "Perceived control" and "ease of information
access" were determined to be factors in how comfortable a person
feels with privacy aspects of using Facebook. Anecdotally, people seem
far more likely to run into trouble with privacy on Facebook than on
Twitter; in the UK for example, there have been several convictions
resulting from criminal evidence posted by
offenders~\cite{Travis2011}. There is also some suggestion that
privacy settings, despite correct use by users, may be overcome
relatively easily, making privacy relatively
illusory~\cite{Zheleva:2009:JJI:1526709.1526781}.

LinkedIn capitalises on its status as a permission-requiring
social-graph medium in that it presents itself as a professional
introductions service. The site encourages strictly professional
information to be shared, and tends to attract older
professionals~\cite{Skeels:2009:SNC:1531674.1531689}. In their review
of social and professional media in the workplace context, Skeels and
Grudin~\cite{Skeels:2009:SNC:1531674.1531689} find tension around
privacy in social media use. However, they also report rapid adoption,
as numerous benefits are found. They found, as will be echoed in our
own findings, that users post more frequently to Facebook than
LinkedIn, demonstrating a particular degree of attraction to that
medium. Though users report great value to LinkedIn, they do not seem
driven to visit the site so often. Our own investigation supports this
observation; reasons for this will be explored later in the
paper. LinkedIn usage seems limited to the professional context,
whereas Twitter seems broadly undifferentiated in that regard, and a
certain amount of Facebook usage will tend to be professional; for
example, Facebook allows businesses to create pages on which they may
publicise themselves.

Although Facebook is the most popular UGM, having over twice as many
users as Twitter, its nearest competitor, the number of connections a
person may have is limited to real-life acquaintances, at least in
principle, and therefore has a practical upper limit of around a few
hundred, with 90\% of users having fewer than 500 Facebook
friends~\cite{Facebook2011}. Interest-graph media users may follow
many more people than they can reasonably keep up with. This can be as
many as several hundred thousand, though Twitter may impose a limit,
and "following" may have other purposes than reading what that person
has to say. This makes microblogging more of a focus of interest when
it comes to information overload. Ramage \emph{et
  al}~\cite{ramage10microblogs} carried out a small scale web-based
survey of 56 Twitter users from within Microsoft.  The focus was on
understanding following behaviour, i.e., how users decide who to
follow, why they follow them, and why they unfollow people. Even
though the focus of this survey was different, its results provide
some useful insights into information overload resulting from
Twitter. For instance, the authors report that ``too many posts in
general'' was the most common reason for a person to be removed from
the user's Twitter stream, i.e., un-followed.  Similarly, with respect
to deciding who to follow on Twitter, the study found that ``...first,
new users have difficulty discovering feeds worth subscribing
to. Later, they have too much content in their feeds, and lose the
most interesting/relevant posts in a stream of thousands of posts of
lesser utility.'' The responses also indicate the need for better
filtering of tweets to more closely match the person's interests, due
to the fact that many Twitter users tend to cover diverse sets of
topics. In that respect, respondents flagged ``too much
status/personal info'' and ``too much content outside my interest
set'' as reasons for un-following Twitter users. The first ties in
with Naaman et al's suggestion that "meformers" may attract fewer
followers, being less likely to be a source of general interest. Also,
in an open-ended question about how the Twitter interface could be
improved, 30\% of the respondents wanted a filtering functionality,
presenting their feeds by user, topic and context.

Our research is a UK-based exploration of UGM use. Questions relating
to usage patterns reflect the topics explored in the Pew Research
survey~\cite{SNS-Pew-11a}, though our sample differs from the Pew
sample in that it tends to be UK-based, and skewed toward the
internet-savvy. In a rapidly changing field, we provide a current
snapshot of UGM use, as well as looking ahead to emerging issues. In
particular, we investigate issues relating to input volume and
overload, so our sample preferentially features internet users, to
increase input from those likely to have encountered those kinds of
issues. We cover similar topics to Ramage et al \emph{et
  al}~\cite{ramage10microblogs} with regards to input volume and
overload, except that our sample size is larger and more diverse,
comprising 587 users recruited via university contacts rather than
being employees within one organisation.


\section{Survey Outline, Aims and Participants}\label{sec:survey-overview}

About half of the questions in our survey related to how people
currently use and engage with UGM, whereas the rest try to gain an
understanding of whether people are experiencing information overload
and from which kinds of UGM. The survey ran over a period of 30 days,
between 29 March 2011 and 29 April 2011.  Since the survey is
targetting user-generated media, such as Social Networking Sites
(SNS), Professional Networking Sites (PNS), and micro-blogs, it was
advertised via Facebook, LinkedIn, and Twitter, as the most widely
used sites of these three kinds of UGM.  The survey was publicised
through several prolific Twitter users, tweeting in their professional
capacity in computer science, and with between 1200 and 4000 followers
each. On Facebook, an open public event was created around the survey
which attracted around 250 participants (although it is not known how
many of them actually completed the survey). On LinkedIn, the survey
was publicised on several professional interest groups, namely Digital
Humanities, Natural Language Processing, Semantic Web Research, Text
Analytics, and User Modelling.  In addition, participants were
recruited via a Sheffield University volunteers mailing list, which is
received by students and academic staff. We were particularly
interested in recruiting a large percentage of young ``digital
natives'', due to their proficiency in and heavy use of social media.

Our questionnaire divides UGM into 5 main kinds:

\begin{itemize}
  \item Social Networking Sites (SNS), e.g. Facebook, MySpace, Bebo;
  \item Professional Networking Sites (PNS), e.g., LinkedIn;
  \item Microblogging sites, and specifically Twitter. While we
    included several social and professional networking sites in the
    survey, to widen its validity, for microblogging we focused
    exclusively on Twitter. This is due to it having the biggest
    growth in data volumes and users amongst all UGM sites;
  \item Social news, e.g. Reddit, Digg, StumbleUpon;
  \item Blogs, e.g., Blogger.com, Wordpress.
\end{itemize} 

The questionnaire begins with demographic questions. It then asks
about the devices respondents use with which they might access the
internet. We ask which social media respondents use. We then ask how
frequently respondents read and post on each type of social media, how
many connections they have on those media where connections are
relevant, their perception of the number of posts they receive
(whether they perceive it to be too many, about right or not enough,
hereafter referred to as volume), the proportion of posts they find
interesting (hereafter referred to as interestingness) and their
ability to keep up with reading all their posts (hereafter referred to
as overload). We ask about reasons for not using social media and
competence to use media should they choose to do so. We ask about use
of mobile phones to access media. We also ask some questions
specifically about Twitter use; to what purpose do respondents put
Twitter. We ask some questions about overload in Twitter. A final
comments field invites criticisms of social media access tools.

\subsection{Participant Demographics} 

We collected 587 complete responses and 228 incomplete ones. The results reported
below are based on the completed answers only.

Our sampling method has impacted on some aspects of the sample. By
gathering data through the university email service and professional
social media presence, the sample contains more students and people
working in related fields than might otherwise be expected. This
contact-based approach to sampling is similar to that used by Pollet
\emph{et al} \cite{Pollet-et-al-2011}, and resulted in similarly
student--biased demographics. 548 of the respondents (93\%) were users
of at least one kind of social media site. This very high percentage
is an artefact of the fact that social media were used as the primary
recruitment mechanism. The figure amongst adults in general, according
to a recent Pew Research Centre survey\cite{SNS-Pew-11a} in America,
is now approaching 60\% of all internet users.

42\% of social media users in our sample are under 25, which is likely
to be a result of the large proportion of student volunteers
(46\%). (Only 2.6\% of the UK population are full-time
students~\cite{HESA2011} and 31\% are under 25~\cite{NatStats2010}).
61\% are female, which is similar to the 56\% reported in
\cite{SNS-Pew-11a}. 94\% of respondents have a laptop computer, 64.5\%
have a smart phone, 8\% have an internet-enabled table, and only 48\%
have a desktop computer.

With respect to country of origin, 493 (84\%) of the participants are from the UK, 
48 (8\%) -- from other European countries, 25 -- from North America (USA and Canada),
19 -- from Asia (primarily India), 1 from Australia and 1 from Africa.

In the ensuing analysis, where it is possible that the bias in our
sample might impact significantly on findings, this is
discussed. Results may be presented treating students and non-students
separately, or above- and below-25s.


\section{General Findings on Use of User-Generated Media}\label{sec:survey-findings}

This section outlines analysis of responses to our questions, focusing
on questions relating to general usage. Each question is taken in
turn. Questions relating to volume, interestingness and overload will
be addressed in the following section in more depth, being of
particular relevance to our research.

\subsection{Uptake}

Amongst respondents, SNS such as Facebook are most widely used 
(94\% of all respondents), followed by PNS (predominenly LinkedIn) (39\%), 
Twitter (37\%), 
blogs (29\%), and social news (11\%). SNS usage percentage is 
comparable to the 92\% reported by the 
Pew Research study in the USA\cite{SNS-Pew-11a}.  
In contrast, reported usage of PNS and Twitter
in our study is significantly higher (39/37 \% vs 18/13 \%), most likely
due to us using these media to recruit participants (since we 
needed a sufficiently large number of users of these media, in 
order to study information overload issues). 


With respect to the extent to which respondents engage with more than
one medium, social networking media seem to occupy a default status;
of those responding that they use PNS, microblogging, social news
sites and blogs, on average 92.75\% also use SNS. Microblogging comes
in next, with 53.44\% of users of other media on average also using
microblogging. Next, PNS at 48.68\%, then blogs at 42.24\% and social
news at 15.93\%. In particular, social news users are most likely to
blog (53.33\% of social news users also blog) and bloggers are most
likely to use social news (19.88\%), bloggers are also most likely to
microblog (62.11\%), microbloggers are most likely to use PNS
(56.56\%) and social news users are most likely to use SNS (98.33\%).

\begin{figure}[]
	\centering
	\caption{Percentage of People Using Medium by Age Group}
	\includegraphics{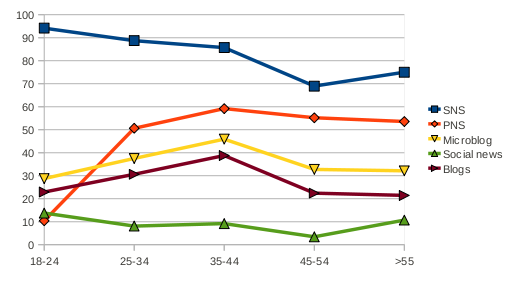}
	\label{fig-age-medium}
\end{figure}

As shown in Figure~\ref{fig-age-medium}, choice of UGM varies across
different age groups. SNS users tend to be significantly younger
(p\textless 0.001, as determined using a two-tailed Spearman's rank
correlation coefficient, as our age data is ordinal) and PNS users
tend to be older (p\textless 0.001). Differences for other media are less
pronounced. Those not using UGM are more likely to be older
(p\textless 0.001). Ramage's study \cite{ramage10microblogs} found that 75\% of
Twitter users were between the ages of 26 and 45, which supports a
mid-range age group for that medium. Significantly different patterns
emerge (p\textless 0.001, Phi/Cramer's V used for nominal media choice data) in
choice of media combinations when we split respondents into those
younger than 25 and those older than 25. 84\% of those under 25 are
using SNS but not PNS, as opposed to 34\% of those older than 25. In
the under 25s, using PNS but not microblogging is very rare (4\%) but
25\% of older respondents do this. 63\% of under 25s use only SNS, as
opposed to 26\% of over 25s. Older people are more likely to use SNS
and PNS but not microblogging (23\% as opposed to 4\%). Older people
are more likely to use all three of SNS, PNS and microblogging (26\%
as opposed to 7\% of under 25s).

Students are significantly less likely to use PNS (p\textless 0.001,
Phi/Cramer's V), and rather less likely to use microblogging (p\textless 0.005)
than non-students. Intuitively, students would most likely be less
drawn to professional networking having generally not yet commenced
their professional life. This is also reflected in significant
differences (p\textless 0.001) in patterns of media choice; among employed
people, 27\% use all three of SNS, PNS and microblogging, as opposed
to 7\% of students. 23\% of employed people use SNS and PNS but don't
tweet, as opposed to 7\% of students. 25\% of employed people use only
SNS, whereas for students the figure is 59\%. 78\% of students use SNS
but not PNS, as opposed to 34\% of employed people. 66\% of students
use SNS but not microblogging, as opposed to 48\% of employed
people. 26\% of employed people use PNS but not microblogging, whereas
for students it is 7\%. These findings broadly reflect the findings on
age; note that the students in our sample are younger (p\textless 0.001), with
79\% being under 25, whereas in the rest of the sample, 91\% are over
25. A sample containing a greater proportion of young people who are
not students may have produced different findings.

Of female respondents, 89\% use SNS, 31\% use PNS, 29\% use
microblogging, 8\% use social news sites and 22\% use blogs. Of male
respondents, 85\% use SNS, 44\% use PNS, 43\% use microblogging, 13\%
use social news sites and 35\% use blogs. Phi and Cramer's V tests
show that the difference between genders in their uptake of these
media is significantly different for microblogging at the 0.001 level
and for PNS and blogs at the 0.005 level, showing in all these cases
that males are more likely to use the medium, whereas for SNS, the
genders are more similar. Males appear also to be more likely to use
social news sites, but due to the smaller overall number of users, a
statistically significant result was not obtained. Ramage's study
\cite{ramage10microblogs} found that 65\% of Twitter users were male,
which is a similar result to ours. Men and women show significantly
different patterns in their choice of media (p\textless 0.001, Phi/Cramer's
V). Men are more likely to use all three of the media of SNS, PNS and
microblogging (24\% of male respondents as opposed to 14\% of female
respondents). Women are more likely to use SNS only (47\% as opposed
to 32\%). 61\% of female respondents use SNS but not PNS, as opposed
to 44\% of male respondents. 62\% of female respondents use SNS but
not microblogging, as opposed to 49\% of male respondents.

Respondents cover a broad range of ethnicities, ranging from white
British (373 respondents) to black Caribbean (1 respondent). Due to
the variation in numbers of respondents of each ethnicity, groups were
combined into British and non-British for the purposes of observing
differences between groups, as this yielded the largest
classes. Results show no significant difference in the uptake of SNS,
microblogging and social news; however, non-Brits are significantly
more likely to use PNS (p\textless 0.001 Phi/Cramer's V) and blogs
(p=0.002). More non-Brits also use SNS (90\% vs. 63\%); however the
difference was not statistically significant. There are marked
differences in the choices Brits and non-Brits make in their media
combinations (p\textless 0.001, Phi/Cramer's V). 29\% of non-British
respondents use all three of SNS, PNS and microblogging, as opposed to
14\% of Brits. 36\% of non-British respondents use SNS but not PNS, as
opposed to 63\% of British respondents. 27\% of non-British
respondents use PNS but not microblogging, as opposed to 12\% of
British respondents. Generally speaking, non-Brits seem to have more
interest in UGM in general and PNS in particular, and where only one
of microblogging and PNS is used, for Brits it will tend to be
microblogging and for non-Brits it will tend to be PNS. Since we don't
have data about where respondents are living, it is unclear if this
difference is due to cultural differences or if non-British
respondents living in this country are perhaps using UGM to keep in
touch with friends back home. However, the most likely explanation is
that the difference arises primarily from age since our non-British
respondents are significantly older (p\textless 0.001) and the non-British
contingent didn't contain so many students.

With respect to employment status, 36\% of all microbloggers in our
sample are students (compared to 46\% in the overall sample), 60\% are
employed or self-employed, and 4\% are currently not employed
(primarily mothers on leave). 211 students in our sample use
professional sites like LinkedIn. In contrast, 163 of the 292 employed
and self-employed respondents of our survey use PNS.




\subsection{Connection Number}

Next, we looked into the number of connections that people have on SNS, PNS, and microblogs. For these three kinds of 
media, connections tend to mean somewhat different things:
\begin{itemize}
  \item On SNS, such as Facebook, people tend to be connected primarily to friends and acquiantances. In addition, as reported by Facebook \cite{FacebookStats2011}, the average user is following up to 80 community pages, groups and events, but these were excluded from our study. 
  \item PNS connections reflect the size of a user's professional network. Users also tend to participate in professional groups around specific interests, e.g. text analysis, and online activity in those groups (e.g. posting articles or jobs, commenting) could lead new professional connections, where the link is established purely due to this virtual communication;
  \item On Twitter people and organisations are followed, primarily due to shared interests, i.e. following/follower networks form the so called \emph{interest graph}. In many cases, the user has not met his connections face-to-face.    
\end{itemize}


\begin{figure}[]
	\centering
	\caption{Number of Connections by Percentage of Users}
	\includegraphics{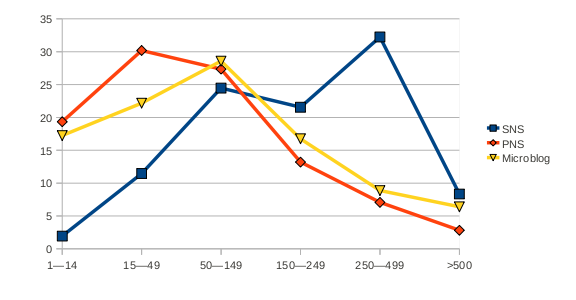}
	\label{fig-connections}
\end{figure}

\begin{figure}[]
	\centering
	\caption{Number of SNS Connections by Number of Users}
	\includegraphics{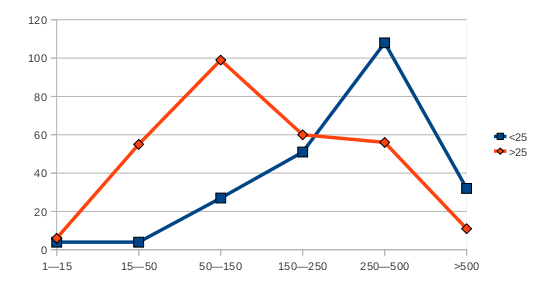}
	\label{fig-sns-connections}
\end{figure}

SNS appears to be the medium that attracts the largest networks, see Figure~\ref{fig-connections}. In our sample, 41\% of SNS users have more than 250 connections, a similar result to the average of 229 reported by Pew Internet \cite{SNS-Pew-11a} and the 180.42 average from the study by Pollet \emph{et al} \cite{Pollet-et-al-2011}. The graph peaks at 50--149 connections and again at 250--499 connections. Breaking down the data into two groups by age helps to explain this; Figure~\ref{fig-sns-connections} shows that for the under 25s, the most usual network size is 250--499 whereas for the over 25s, it is 50--149. For the under 25s, (44\% of SNS users in our sample), SNS are the user-generated media of choice, with most users (62\%) having in excess of 250 connections. Younger people have significantly more connections on SNS than older users (p\textless 0.001, Spearman's), a result more striking given that younger people have had less time to collect acquaintances in their lives. In contrast, SNS users who are 35 or older most commonly have 50-150 friends in their network (36\%), with 68\% having 150 connections or fewer. Given that 64\% of them also use a PNS, 42\% use a micro-blogging service and 36\% are also bloggers, our findings suggest that older users tend to have separate online identities and networks maintained through different UGM sites.

For PNS, 105 (49.5\%) of our respondents have 49 or fewer connections. From these, 16 (15\%) are students (the rest are employed or self-employed) and 19 (18\%) are under the age of 25. In contrast, there are only 3 students and only 6 under-25s (5.6\%) amongst the 107 respondents with more than 50 connections. In our sample, 25-34 year olds are just as likely to have under 50 connections as they are to have more (41 vs 40), whereas 60\% of over 35s have more than 50 professional connections. These findings are in themselves not surprising, since professional networking sites are designed exactly for employed users, whose professional networks grow as they meet more people throughout their career (hence the connections increasing with age).

The average number of connections on micro-blogging sites is 138, with the largest number (58) in the region of 50--149 connections. However, unlike PNS, here only 40\% of users have fewer than 50 connections and the number of users with 500+ connections is double that for PNS. Two years ago, a Twitter survey \cite{Cheng2009} reported that 92\% of Twitter users follow less than 100 people. However, since then the number of Twitter accounts has grown from 11.5 million to more than 200 million and consequently, so has the average number of people and organisations followed. Age has little impact on the number of connections respondents have on microblogs.

There is little evidence of a gender difference in the number of connections respondents have on SNS, PNS and microblogs. As noted earlier, non-Brits are more likely to use PNS; however even within PNS users, non-Brits are significantly more likely to have more connections (p=0.001). On other media, number of connections are similar for Brits and non-Brits. Mobile-users have slightly more connections on SNS (p=0.001), though not on PNS or microblogs. It might seem that this could be accounted for if mobile users tend to be younger, since younger people have more connections on SNS; however no evidence was found that mobile users tend to be younger.


\subsection{Frequency of Access}

\begin{figure}[]
	\centering
	\caption{Frequency of Reading}
	\includegraphics{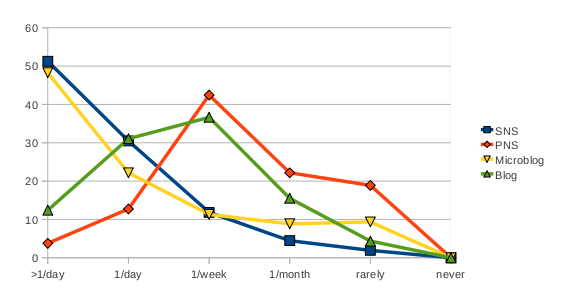}
	\label{fig-read-freq}
\end{figure}

With respect to frequency of media use, there are significant 
differences between SNS and microblog 
engagement on one hand, 
and PNS (LinkedIn) on the other (with blogging falling in the middle). 
Figure~\ref{fig-read-freq} shows the frequency with which users of each medium read posts. Note that percentages are calculated within the medium, not across the entire sample. The graph shows that almost 82\% of SNS and 71\% of microblog users read posts daily or more, whereas only 44\% of bloggers do so, falling sharply to below 17\% for PNS users. Blogs and PNS show a peak around reading once per week. Similar results are reported 
also by Pew Internet \cite{SNS-Pew-11a} and reflect the predominently 
real-time, status-like content on Facebook and Twitter. However, 
it is also possible that the higher volumes of content shared daily
through these kinds of UGM, might lead to many people feeling the 
need to check their streams at least twice daily,
in order to stay on top of important, real-time information. 


\begin{figure}[]
	\centering
	\caption{Frequency of Posting}
	\includegraphics{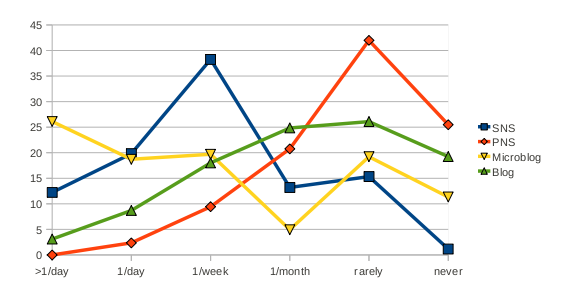}
	\label{fig-post-freq}
\end{figure}

When it comes to frequency of posting, the picture is more complex. As Figure~\ref{fig-post-freq} shows, microblogging is the only medium where users are most likely to post more than once per day, with 26\% of microblog users doing so. SNS shows the most similar posting pattern to microblogging, with something of a bimodal pattern emerging as users are divided into more or less frequent posters, lending credence to the anecdotal "lurker" pattern; a user who rarely posts but reads regularly. Blog and PNS use most commonly post "hardly ever" (26.09\% and 41.98\% respectively), with very small numbers posting once per day or more. Younger SNS users read more (p\textless 0.001, Spearman's) and post more (p\textless 0.001). Ramage's study \cite{ramage10microblogs} found that 67\% of Twitter users were very active consumers of information, reading posts several times a day. 37\% posted more than once per day, and 54\% posted with frequency between once a day and once a month. This is not dissimilar to our findings.

There is little evidence of a gender difference in frequency of posting and reading, except that males are slightly more likely to post to microblogs (p\textless 0.005, Phi/Cramer's V). Brits post more on SNS (p=0.001). Non-Brits do not read or post more on PNS despite the greater uptake and number of connections, suggesting the difference may be accounted for more by prevalence in certain communities than a greater cultural enthusiasm for the medium.

In general, users who frequently engage with one medium are somewhat more likely to engage frequently with the others also. Reading SNS correlates significantly with reading microblogging sites (p=0.001). Reading microblogging sites also correlates significantly with reading blogs (p\textless 0.001). In other respects, though positive, correlations are not significant. Posting, however, shows a more consistent relationship across media. Frequent posters on SNS are likely to be frequent posters on PNS (p\textless 0.001), microblogs (p\textless 0.001) and blogs (p=0.002). PNS posters are also bloggers (p=0.002), and microbloggers are also bloggers (p=0.001). There may be a slight suggestion here that reading behaviour is governed more by preferred media, whereas posters will post on a wide range of media. Number of connections over SNS, PNS and microblogs is in all cases correlated to a highly significant degree.

358 respondents of 515 (61\%) indicated that they have an internet-enabled mobile phone. It might be theorised that users with internet-enabled phones would access UGM more frequently since it is convenient for them to do so. However, no significant result was found to this effect.

\subsection{Mobile Devices and Social Media}

\begin{figure}[]
	\centering
	\caption{Frequency of Use of Mobile Devices for Different Purposes}
	\includegraphics{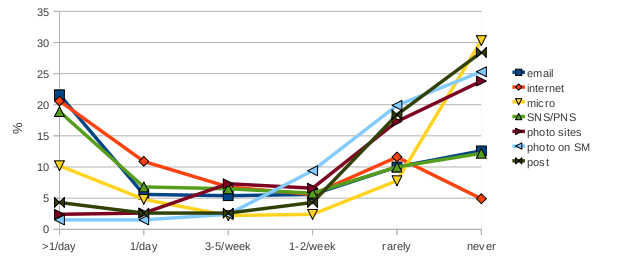}
	\label{fig-sms-usages}
\end{figure}

A number of questions specifically targeted those with web-enabled phones. Respondents were asked about the frequency with which they use their web-enabled phone to check email, read the internet, read microblogs, read SNS and PNS sites, use photo and video sharing sites, take and share photos on social media and share web content on social media. Responses to these questions show a bimodal distribution, with users typically either doing these things often or else rarely or never. In particular, email (27.2\%), reading sites such as news on the internet (31.5\%) and using social media (25.7\%) are those tasks that a significant percentage of users do once per day or more on their web-enabled phone. 15\% of respondents use their phone for microblogging once per day or more. Other tasks were not popular usages of these devices. Figure~\ref{fig-sms-usages} illustrates responses to these questions.

\subsection{Not Using Social Media}

The responses from the 7\% (39) of survey participants who do not use
any UGM indicate that 61\% of them are female, 44\% are under the age
of 35, and 72\% hold one or more university degrees.

When asked why they do not use Twitter or Facebook, the majority
response (25.45\%) is that they are not interested in doing so. Those
that provide more specific information commonly cite content value
(18.18\%) and privacy concerns (16.36\%) as their reasons for not
using the media. Lifestyle aspects such as preference for face-to-face
interaction or not often using the computer or preferring not to
account for a further 18.18\% of reasons given. 12.72\% don't have
time to use the media and 7.27\% are concerned about bullying and
abuse. One user (1.81\%) specifically cites trouble with finding the
relevant content. For PNS, the picture is slightly different, with the
majority reason for not using being a perception that it is irrelevant
or that they don't know what it is. This perhaps reflects the large
number of students in the sample.

76\% of respondents who say they do not use social media say they know
enough about UGM, but do not wish to use it. In other words, for our
participant sample, non-use of social media is not due to being older
or lacking knowledge/expertise.


There is a colourful variety in the ways people express their lack of
interest in the content provided on microblogging sites. "Worthless
drivel", "self-important ramblings", "trivial rubbish", "not
interested in what celebrities had for breakfast", "narcissistic" and
"people shout about their own lives too much" are some of the comments
people made, and largely reflect the tone of comments with regards to
not using microblogging.

For SNS, some similar comments appear, such as "indifferent to other
people's trivia". However, comments are often more personal,
reflecting a complexity in fitting Facebook into their lives. One
user's boyfriend wanted them to stop using the site. Another spends
"all day at work on the computer" and doesn't want to spend leisure
time in front of the screen. One user thinks "it's a bullying tool";
another thinks it is "too fraught with risks". One user perceives that
their profession requires them to remain private. Other similar
comments reflect a thoughtful, more involved relationship with not
using social networking.

\section{Findings Related to Volume, Interestingness and Overload}\label{sec:survey-findings-overload}

This section outlines findings related to volume of posts/comments
received, the level of interest respondents have in these posts and
whether they perceive that they are receiving too many. Each of these
will be taken in turn. Then the issues will be explored in the broader
context of responses to other questions for each medium in
turn. Finally, respondents' overall comments will be discussed with
particular attention to volume, interestingness and overload.

\subsection{Perceptions of Number of Posts Received (Volume)}

For SNS, the majority perceive the number of posts they receive to be
about right for them (72\%), and of the remainder, slightly more
perceive that they receive too many posts rather than too few (21.3\%
as opposed to 6.6\%).  For PNS, the picture changes slightly. Again,
most (68.4\%) perceive that they receive the right number of posts,
but of the remainder, 23\%, as opposed to 8.6\%, would rather receive
more. This ties in with the earlier finding that PNS users post less
(88.3\% of users posting monthly or less on PNS, whereas 70.3\% of SNS
users post at least weekly).  Microblogging users again mostly
perceive that they receive the right number of posts for them
(58.3\%), but this time, more users perceive that they receive too
many posts (33.9\%), with only 7.8\% receiving too few. Respondents
who use both SNS and PNS are significantly more likely to percieve
that they receive too many posts on SNS than PNS (p\textless 0.001, determined
using Wilcoxon signed rank test, a non-parametric paired test). Those
who use microblogs and PNS are significantly more likely to perceive
that they receive too many posts on microblogs (p\textless 0.001).

\begin{figure}[]
	\centering
	\caption{Volume}
	\includegraphics{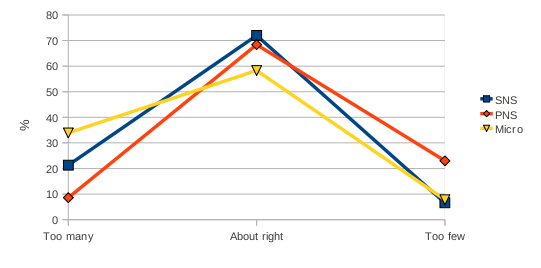}
	\label{fig-sms-received}
\end{figure}

Respondents who perceive that they receive too many posts on social
media are significantly more likely to have the same perception on PNS
(p=0.001) and microblogging sites (p\textless 0.001). However, perception of
number of posts received on PNS does not correlate quite so well with
perception of number of posts received on microblogging sites.

There was no evidence to suggest that gender, occupation (student or
not) or whether a respondent uses an internet-enabled mobile phone
have any influence on users' perception of volume. However younger
users were less likely to perceive that they receive too many posts on
SNS (p=0.001, Spearman's) and microblogging sites (p=0.006). No age
correlation was found with perception of volume on PNS.

\subsection{Interest Level in Posts Received (Interestingness)}

For SNS, the majority, 35.7\%, say that the posts they receive are
half interesting, half boring, with slightly more of the remainder
(37.5\% as opposed to 26.8\%) saying that they find more of the posts
uninteresting than interesting. With regards to the proportion of
interesting posts received, opinion is more spread on PNS. 18.4\% say
they find the posts half interesting, half boring, with 58\% more
bored, and 23.6\% less bored. Generally speaking, PNS are perceived to
be more boring. Of microblog users, 36\% find the posts they receive
half interesting, half boring, with 31\% generally more interested
than that, and 33\% less interested. This gives microblogging a
slightly higher interest level than social or professional networking
media, perhaps because Twitter is based on the interest graph (linking
those with similar interests) rather than the social graph of
describing existing relationships. In terms of interest level, social
news sites and blogs take the prize, as the majority response for both
media was that most posts are of interest to them (41.7\% for social
news sites and 42.9\% for blogs). Comparing the responses of those
respondents who use both media, blogs were found significantly more
interesting than SNS (p\textless 0.001, Wilcoxon) and significantly more
interesting than PNS (p\textless 0.001) and blogs were significantly more
interesting than microblogs (p=0.001). Other results were less marked,
though SNS is perceived as somewhat more interesting than PNS by those
users who use both (p\textless 0.05).

\begin{figure}[]
	\centering
	\caption{Interestingness}
	\includegraphics{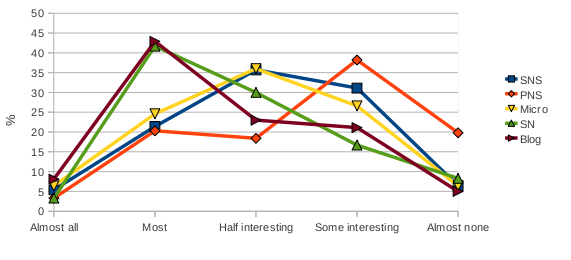}
	\label{fig-sms-interest}
\end{figure}

Generally speaking, respondents who perceive a high degree of interest
in one social medium are likely to perceive more interest in others;
all correlations were positive, and all were significant, except for
interest levels in social news, which did not correlate so well with
the other media. (SNS, PNS, p\textless 0.001; SNS, micro, p=0.008; SNS, blogs,
p=0.005; PNS, micro, p=0.006; PNS, blogs, p=0.005; micro, blogs,
p\textless 0.001).

Respondents with internet-enabled mobile phones are slightly more
likely to find PNS interesting than those with no mobile internet
access (p=0.01, Phi/Cramer's V). There was no evidence to suggest that
gender, occupation (student or not) or whether a respondent uses an
internet-enabled mobile phone have any influence on the extent to
which the media are perceived as interesting. Younger people find
posts more interesting on SNS (p\textless 0.001) and PNS (p=0.002), with less
significant correlations to a similar effect on microblogs and social
news. Perhaps older people have read it all before?

\subsection{Ability to Keep Up with Posts Received (Overload)}

Most SNS users are able to keep up with reading the posts they receive
(82.3\% say they always or mostly read and respond to others' posts)
with the remainder missing important posts or failing to keep up with
the volumes received.  73.4\% of PNS users mostly or always keep on
top of the posts they receive, with 26.6\% missing important posts or
failing to keep on top of posts received. Notice that despite fewer
PNS users perceiving that they are overloaded, they nonetheless are
less likely to keep up. Generally speaking, microbloggers are able to
keep on top of the posts they receive (55.6\% mostly or always reading
and replying to posts received); however at 44.4\%, microbloggers are
those most often failing to keep up. Social news users and blog-users
show a similar profile to microbloggers in terms of their ability to
keep on top of their reading, perhaps reflecting that these sites all
focus on the interest graph. Those respondents who use both microblog
sites and SNS are significantly more likely to keep up with the SNS
site (p\textless 0.001, Wilcoxon); those who use social news sites and SNS are
significantly more likely to keep up with the SNS (p=0.001); likewise
those who use blogs and SNS are significantly more likely to keep up
with the SNS (p\textless 0.001).

\begin{figure}[]
	\centering
	\caption{Overload}
	\includegraphics{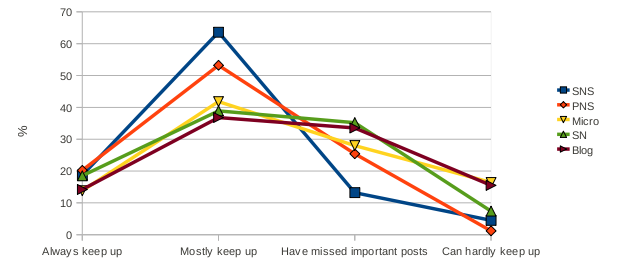}
	\label{fig-sms-upkeep}
\end{figure}

In the general case, respondents who claim to keep up with their
reading on one medium also keep up with their reading on other media,
though the relationship is not significant in all cases (SNS, PNS,
p\textless 0.001; PNS, micro, p\textless 0.001; PNS, blogs, p\textless 0.001; micro, SN, p\textless 0.001;
micro, blogs, p\textless 0.001; SN, blogs, p=0.009).

There was no evidence to suggest that gender, occupation (student or
not) or whether a respondent uses an internet-enabled mobile phone
have any influence on overload. Younger people say that they keep up
better with their SNS reading (p\textless 0.001), with no significant
correlations found between age and ability to keep up with reading on
other media.

\subsection{Facebook and Social Networking Analysis}

Frequency of reading and posting are highly correlated (p\textless 0.001,
Spearman's). Those who read and post frequently are more likely to
find posts interesting (p\textless 0.001, p\textless 0.001) and less likely to
experience overload (p\textless 0.001, p=000), and those who read frequently
are less likely to perceive themselves to be receiving too great a
volume of posts (p\textless 0.001).

Those with more connections on Facebook and other SNS post more
(p\textless 0.001) and read more (p\textless 0.001). They are also significantly more
likely to claim interest in a greater proportion of the posts they
receive (p\textless 0.001). Whether a user has many connections or few has
little impact on perception of volume. In other words, the number of
posts received in absolute terms does not affect whether a respondent
perceives themselves to be receiving the right number of posts (since
receiving more posts inevitably results from having more
connections). This suggests that those who are more invested in the
medium, having more connections and reading and posting more often,
are the ones who experience less overload, not those with fewer posts
to read. In fact, a weak correlation was found (p>0.05) to the effect
that those with fewer connections are more likely to experience
overload.

Those who perceive that they receive too great a volume of posts are
less likely to find the posts interesting (p\textless 0.001) and more likely to
experience overload (p\textless 0.001). Those who find the posts interesting
are less likely to experience overload (p\textless 0.001).

\subsection{LinkedIn and Professional Networking Analysis}

As with social networking sites, professional networking site users
who read more also post more (p\textless 0.001, Spearman's), have more
connections (p\textless 0.001), are less likely to perceive that they are
receiving too great a volume of comments (p=0.001) and find more of
the posts interesting (p\textless 0.001). Users who post more have more
connections (p\textless 0.001 and find more of the posts interesting (p\textless 0.001);
however there is no correlation between posting frequently and
perception of volume, unlike in SNS where frequent posters are less
likely to perceive that they receive too great a volume of posts. No
correlation was found between frequency of reading and posting and
overload.

Unlike with social networking, where no relationship was found, for
PNS, well-connected users are less likely to perceive that they
receive too great a volume of posts (p\textless 0.01). Well-connected users, on
the other hand, are not especially more likely to find the posts they
receive interesting. The fact that users have a greater appetite for
posts on professional media may reflect that people are posting less
on these media; the most common frequency of posting on social media
is "weekly" (38.3\%), with most users posting at least weekly
(70.3\%), whereas for professional media, it is "hardly ever" (42\%),
with the majority of users posting hardly ever or never
(67.5\%). There are quite simply fewer posts on professional
media. Having more connections does not correlate with overload or
lack thereof.

It is interesting that well connected users do not find the posts they
receive more interesting than less well connected users. This suggests
that interest in others' posts is not the driving factor in investment
in the medium. Perhaps this reflects that professional networking is
something one does with the aim of professional advancement, so users
may be tolerant of a lower level of gratification. It may be that
users connect with others on PNS for the purposes of documenting and
extending their network rather than as a means to access certain
peoples' posts; in other words, PNS may not primarily be about posts
and updates. The relationship between perceiving excessive volume in
the number of posts received and overload is weaker in professional
media as opposed to social, suggesting unclear desires or a more
complex relationship with regards to posts received.

\subsection{Twitter and Other Status Update Media Analysis}

As with social and professional media, reading frequently correlates
with posting frequently (p\textless 0.001), having more connections (p\textless 0.001),
finding more of those posts interesting (p\textless 0.001) and not experiencing
overload (p=0.002). Posting frequently correlates with having more
connections (p\textless 0.001), finding more posts interesting (p\textless 0.001) and
not being overloaded (p=0.005). Having more connections correlates
with finding more posts interesting (p=0.001) though not with (lack
of) overload, as for PNS. Predictably, those who perceive that they
are receiving too great a volume of posts are also those who are more
overloaded (p\textless 0.001). Those who find the posts more interesting are
less overloaded (p\textless 0.001).

We included a number of questions specifically relating to overload on
Twitter, in order to gain an insight into how people use the
micro-blogging platform. 38\% of tweeters in our study were using it
both in personal and professional capacity, 38\% for personal use only
and 24\% for professional use only, making usage fairly
balanced. 51.5\% use Twitter to follow the status updates of friends,
family and/or celebrities. 48.5\% use it to get professional
information from colleagues. 46.6\% get news updates via twitter and
34.3\% converse with friends. 15.2\% ask questions and get help via
Twitter. Those who use Twitter for personal use are more likely to
follow friends, family and celebrities (63.04\% as opposed to
36.61\%). Those who use Twitter for professional purposes are more
likely to use Twitter to get professional information (73.21\% as
opposed to 42.75\%).

Of Twitter users, 64.9\% have at some time unfollowed a user because
they post too much, with 20.2\% doing this in the last week or "all
the time". 70.4\% have found it difficult to locate the
important/interesting posts amongst the others, with 23.7\% finding
this all the time, and 24.7\% having found this within the last
week. 66.3\% have felt at some time that they receive too many tweets
and can't keep up, with 21.2\% of those feeling this way all the time,
and 24.5\% having felt this way in the last week.

44.1\% have unfollowed a user because they did not post enough
interesting information, with 18.8\% doing so in the last week or all
the time. 55.6\% have felt the need for a tool that filters out
irrelevant posts.

Those who use Twitter in a personal capacity read more (p\textless 0.01,
Phi/Cramer's V), post more (p\textless 0.05), perceive less excess of volume
(p\textless 0.01) and experience less overload (p\textless 0.01) than those who use
Twitter in a professional capacity only. They are somewhat more likely
to un-follow another user because they do not post enough interesting
information than those who use Twitter in a professional capacity only
(p\textless 0.05). They are also slightly more likely to feel the need for a
tool that filters out irrelevant posts (p=0.05).

Those who use Twitter in a professional capacity do not especially
read more or less than those who use Twitter in a personal capacity
only; however they post more (p\textless 0.05), have more connections (p\textless 0.05)
and are more likely to perceive that they receive too great a volume
of posts (p\textless 0.01). They are significantly more likely (p\textless 0.001) to
feel that they receive too many posts and are overloaded than those
who use Twitter in a personal capacity only.

These results echo those found in social and professional media, where
social users have less tolerance for uninteresting posts, and
professional users don't keep up as well with their reading. A picture
also starts to emerge of professional users being perhaps more focused
on attracting attention than giving it; perhaps there are more
broadcasters or self-promoters among professional users. Intuitively
speaking, some successful broadcasters such as known celebrities will
be using Twitter in a professional capacity only, and their followers
will perceive themselves to be using Twitter in a personal capacity,
following those people out of pure interest. This takes the difference
between personal and professional users to its extreme, illustrating
how a personal user may be interest-led, where a professional user may
be focused on projecting an image or attracting fans. It is unclear
how many successful celebrities responded to our survey, but the same
dynamic may play out to a lesser extent among more typical users.

Note however that there are potentially different models of
professional Twitter use, with self-publicisers, who are likely to be
"meformers" [Naaman et al], being only one possibility. Experts such
as professional bloggers may be classic "informers" [Naaman et al],
primarily attracting followers through the dissemination of
interesting information, although research has shown that these are
rarer. Another type of professional use may take the form of workplace
or workgroup cliques, where Twitter is used as the medium of
communication within the group. Further research would be required to
determine the prevalence of these patterns, but the clique pattern
seems more likely to reflect social media usage patterns, and thus
wouldn't affect our findings here other than to weaken the difference
between social and professional use.

Those who use Twitter in both a personal and a professional capacity,
as opposed to those who use Twitter for either professional or
personal purposes but not both, read more (p\textless 0.01) and post more
(p\textless 0.01), have more connections (p\textless 0.001) and are more overloaded
(p\textless 0.05). They are more likely to have un-followed a user because they
post too much (p\textless 0.01) and more likely to have un-followed a user
because they did not post enough interesting information (p\textless 0.05).

\subsection{Social News and Blogs Analysis}

For social news sites, again, posting and reading are correlated,
though less significantly so than for other media (p=0.01), perhaps
due to the smaller number of respondents. Those who read more also
tend to find more of the posts interesting and tend to experience less
overload, though this result was not significant. Those who post more
do not especially tend to keep up better and find more of the posts
interesting, perhaps hinting that the medium attracts a "broadcaster"
user subtype, more focused on disseminating information than receiving
it, though more data would have to be gathered to form any conclusions
regarding this medium.

Bloggers who read more also post significantly more (p\textless 0.001), as for
the other media. They are also significantly more likely to find the
posts they read interesting (p\textless 0.001) and less likely to be overloaded
(p\textless 0.01). Bloggers who post more find posts more interesting (p\textless 0.001)
and experience less overload (p\textless 0.01). Those who find blogs
interesting are less likely to be overloaded (p\textless 0.001).

\subsection{Applications Used, Shortcomings and Desired Functionalities}

A final question asked users to comment on "which applications do you
use to read social media, their main shortcomings, and what an ideal
social media access tool would have in terms of functionalities, which
you're currently missing?" Responses illustrated a varying degree of
technological sophistication among respondents. For example, with
regards to the applications used to access the media, responses range
from "not sure what's meant by applications", "I have already said I
have a laptop and internet-accessible phone", "The internet?" through
to "hootsuite, tweetdeck", "Twitter.com, qTwitter, gwibber", "Twitbird
Pro, Official Twitter App, Osfoora" and other such responses
reflecting an awareness of options for accessing the media. Of 272
respondents to this question, only 80 demonstrated in their response a
clear understanding of what an application is. The others,
potentially, remain limited to the official website or default client
on their device through an unawareness of alternatives.

\begin{figure}[]
	\centering
	\caption{Complaints and Concerns}
	\includegraphics{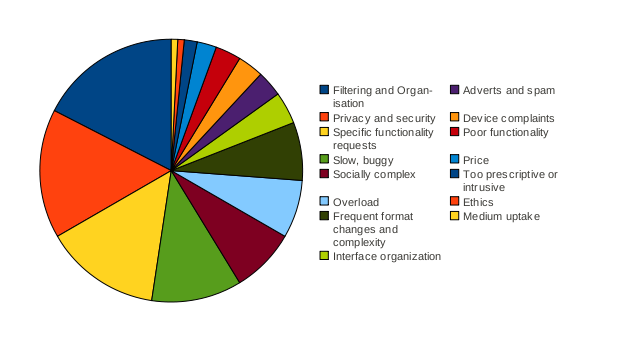}
	\label{fig-comment-pie}
\end{figure}

As a result of a poor understanding of the concept of an application,
responses to this question included comments on media and devices as
well as applications. The responses were interesting nonetheless. 126
respondents gave complaints and described desired functionality, with
the remainder having no opinion or expressing satisfaction with the
functionality already available to them. The most common response,
from 22 respondents, was that better functionality for filtering out
the relevant from the irrelevant posts would be desirable; a further 9
respondents complained of overload in the posts they receive. 15
respondents are concerned about privacy, with a further 5 concerned
about security issues. 14 find their application slow or buggy. 16
users expressed a desire for a specific piece of functionality. These
were generally speaking very varied: for example, three users would
like an application that combines all their media into one; two users
would like to see video chat included and one user would like to see
summarisation functionality. 10 respondents find the media socially or
personally challenging in some way; addictive, demanding, distracting
or leading to tension. Other concerns include price, ethical concerns
and adverts and spam. 9 respondents complained about interface
complexity, with the majority of those specifically citing Facebook's
recent frequent interface changes. Figure~\ref{fig-comment-pie} shows
the proportion of respondents expressing the different concern types
in graphical form.

Some of those respondents who focused on filtering and overload made
statements of the problem; "gets too cluttered with rubbish," "not
optimal for filering og searching [sic]," "the main problem is the
volume of information," "too many useless and time wasting tweets," "I
particularly dislike people who post the same thing many times." Other
respondents stated possible solutions, some vague and others more
specific; "would be good to have a facility that automatically puts
tweets into a folder categorized by specified hashtags," "tool to
categorize topics by importance," "something that conflates identical
posts would be useful," "social sites could do with a digest option,"
"Increased filtering options would be nice," "it would be great if it
could delete irrelevant posts for me." Others perhaps more
optimistically focus on quite advanced desired functionality; "I think
the automatic summarization on public opinions would be great!" and "I
would like some tool to, regardless the twitter account I am using (if
more than one) allow me to have a single but also debriefed glimpse of
my interest."


\section{Discussion}\label{sec:conclusion}

Results generally speaking show a high degree of engagement with
social networking media, and a significant engagement with other
media. The general picture presented by our findings is one of broad
satisfaction, with the majority of the users perceiving that they have
no major complaints regarding the media. Employed users make more use
of professional media than students. Male users are more likely to use
a range of media, with women more likely than men to use social media
only. Younger respondents are more likely to use social media only,
with older respondents using a range of media, though note that our
younger respondents were more likely to be students and therefore have
less interest in professional media. Under 25s have more connections
on social media and older people have fewer. The number of connections
people have on professional media and microblogging sites tends to be
lower. Social media and microblogs are most often read once a day or
more, whereas blogs and professional media are most often read once
per week. Males post more often to microblogs. Blogs and professional
media receive a lower volume of posts. For social media and
microblogs, users may fall into two camps; frequent posters and
infrequent posters.

Most users are satisfied with the amount of information they receive
and are able to keep up with it. Overload appears better related to
the extent of engagement with the media than with quantity of posts
received in absolute terms; users with more connections report less
overload despite their inevitably receiving more posts. Users engage
frequently with the media despite middling-to-low perceptions of the
interest of posts received in some cases; this seems to be accepted as
par for the course. However a significant minority express a desire to
improve the situation through filtering functionality.

It is possible that the questions about Twitter, in particular, "[have
  you] Felt the need for a tool that filters out irrelevant posts",
may have primed the user to give more answers around this topic in the
final comments section. However, although this may have influenced the
number of users making this kind of comment, it doesn't change the
original finding that 55.6\% of respondents indicate that they have
felt the need for such a tool, and the free text section at the end of
the questionnaire provided a valuable forum for users to speak more
freely on the matter. For example, some users talk about wanting a
ranking system, whereas others would like irrelevant posts to be
deleted. Some users talk about quality control where others want the
posts of closer friends to be prioritised. It is clear that
investigations in the area of ranking at summarisation would be
welcomed by a significant number of users.

One of the questions in the survey asks users about their ability to
keep up with the information they receive; for example, have they
missed important/interesting information. Microbloggers complain of
overload to the greatest extent. The majority of users, especially for
SNS and PNS, claim that they mostly keep up and do not miss important
posts; however, this raises the question of how they would know if
they had missed an important post. Information overload leads to the
problem of information overlook; "missing critical information that
has been clouded by information overload" [Lansing Taylor], and the
user will be unaware that they have missed this information, creating
the illusion that they are better informed than they actually are (in
the words of Donald Rumsfeld, they are "unknown unknowns"). The
deception may be compounded by the "recency illusion" []; a user who
has recently become aware of a new concept on social media may feel
pleased to be informed of it in a timely manner, and it may not not
occur to them that the concept may have been around for years or
decades unbeknownst to them. In fact, the extent to which a person
feels satisfied that they are informed relates better to how recently
they have invested time in it, and learned something, whatever it may
be, than it does to the actual extent to which they are informed, a
quantity they cannot possibly know.


A survey such as this can only scratch the surface of this issue,
depending as it does on respondents' own subjective perceptions of the
efficacy of the media in bringing the right information to their
attention. Controlled tests would need to be developed to assess the
extent of the problem in practical terms; however, on a philosophical
level the question answers itself. The amount of information in the
world is vast and ever increasing, and one human being can only ever
hope to scratch the surface of it.

The subject of overload and filtering can be approached from another
angle. The problem facing a Twitter user is twofold; firstly to
optimise the ratio of interesting tweets read to total number of
tweets read (precision) and secondly to maximise the total number of
interesting posts read (recall). Users complaining of requiring a
filtering tool are dissatisfied with their precision, that is to say,
they are having to read too much uninteresting input. But precision
and recall trade off against each other; a user may increase precision
by focusing on, for example, only reliably interesting tweeters, if
they are prepared to sacrifice some recall, i.e. obtain in total a
lower amount of interesting information. A hypothetical user, given a
way of improving their precision (such as a filtering tool) will
simply increase the amount of material consumed to bring precision
back down to their personal tolerance level and benefiting from
improved recall whilst holding precision constant. Provision of
filtering tools isn't really about improving precision, that is to
say, getting the uninteresting posts off people's screens. Filtering
tools are about improving accessible recall.

A difference in professional and personal media usage emerged in the
research, with professional users being less interest-driven, posting
and connecting despite a lower degree of interest in reading what
other users have to say. This raises questions about what captures
people's interest. It seems that when it comes to social media, people
are making more of an effort to keep up, perhaps because connections
are personal on social media, and one wouldn't want to ignore a
friend. However, keeping up isn't purely obligation driven, as
respondents also find social media more consistently interesting. It
seems likely there is a tendency to find people more interesting
purely because they are people we know, and may be talking about
events that affect us or other people that we also know. When it comes
to the interest graph, however, the picture is different. Number of
followers on Twitter follows a Zipf-like pattern
(http://www.hubspot.com/Portals/53/docs/01.10.sot.report.pdf), with a
small number of users having a very high number of followers, and
number of followers having no clear relationship with the number of
followed. Personal relationships force a more even distribution, as
these are people we have to make time for in our lives, but the
interest distribution is decidedly unfair. As soon as we move into
professional usage, our research has shown, usage patterns drift
closer to the interest model, with a disparity between the amount of
attention sought and the amount of attention given. This works well on
media such as Twitter, where personal-usage "receivers" follow
professional-usage "broadcasters"; however on PNS a level of tension
is evident in the response pattern to our questions about the
medium. Somehow, as a status update medium, it falls awkwardly in the
gap between social- and interest-graph media.

Most participants are unaware of the possibility of using alternative
clients to customise their experience of the media. This has
implications for the provision of filtering functionality, as many
users may not be sufficiently technologically sophisticated to
benefit. Technological and social sophistication required by the media
also emerge as themes in the free text section. Filtering,
organisation and overload feature prominently; however privacy,
security and the perceived dangers of the internet also loom
large. Interface complexity also comes up as a concern. There is a
sense that in order to benefit from the media, a level of investment
is required in learning how to use the technology and keeping abreast
of its updates, and that for some people, this may be prohibitive.

\bibliographystyle{alpha}
\bibliography{../../../../../big}

\begin{thebibliography}{KLPM10}

\bibitem[Age11]{HESA2011}
Higher Education~Statistics Agency.
\newblock {Statistics - Students and qualifiers at UK HE institutions}, 2011.
\newblock http://www.hesa.ac.uk/content/view/1897/239/.

\bibitem[Arr06]{Arrington2006}
Michael Arrington.
\newblock {Odeo Releases Twttr}.
\newblock {\em TechCrunch}, 2006.
\newblock {\small {http://techcrunch.com/2006/07/15/is-twttr-interesting/}}.

\bibitem[CE09]{Cheng2009}
Alex Cheng and Mark Evans.
\newblock {An In-Depth Look Inside the Twitter World}.
\newblock {Technical report}, Sysomos Inc., 2009.
\newblock {\small {http://www.sysomos.com/insidetwitter/}}.

\bibitem[Dou07]{Douglas2007}
Nick Douglas.
\newblock {Twitter blows up at SXSW Conference}.
\newblock {\em Gawker.com}, 2007.
\newblock {\small
  {http://gawker.com/243634/twitter-blows-up-at-sxsw-conference?tag=technextbigthing}}.

\bibitem[ES10]{Ehrlich2010}
Kate Ehrlich and N.~Sadat Shami.
\newblock Microblogging inside and outside the workplace.
\newblock In {\em Proceedings of the Fourth International AAAI Conference on
  Weblogs and Social Media}, pages 42--49. AAAI, 2010.

\bibitem[{Fac}11]{FacebookStats2011}
{Facebook}.
\newblock {Statistics}.
\newblock http://www.facebook.com/press/info.php?statistics. Accessed on July
  21st, 2011., 2011.

\bibitem[fNS10]{NatStats2010}
Office for National~Statistics.
\newblock {National Population Projections, 2010-based projections: Principal
  projection - GB population in age groups}, 2010.
\newblock
  http://www.ons.gov.uk/ons/rel/npp/national-population-projections/2010-based-projections/rft-table-a2-1-principal-projection---uk-population-in-age-groups.xls.

\bibitem[HGRP11]{SNS-Pew-11a}
Keith~N. Hampton, Lauren~Sessions Goulet, Lee Rainie, and Kristen Purcell.
\newblock Social networking sites and our lives: How people's trust, personal
  relationships, and civic and political involvement are connected to their use
  of social networking sites and other technologies.
\newblock Technical Report June 16, Pew Research Center's Internet and American
  Life Project,
  http://pewinternet.org/Reports/2011/Technology-and-social-networks.aspx,
  2011.

\bibitem[HT85]{Hiltz:1985:SCC:3894.3895}
Starr~R. Hiltz and Murray Turoff.
\newblock Structuring computer-mediated communication systems to avoid
  information overload.
\newblock {\em Commun. ACM}, 28:680--689, July 1985.

\bibitem[HXLR10]{Hoadley201050}
Christopher~M. Hoadley, Heng Xu, Joey~J. Lee, and Mary~Beth Rosson.
\newblock Privacy as information access and illusory control: The case of the
  facebook news feed privacy outcry.
\newblock {\em Electronic Commerce Research and Applications}, 9(1):50 -- 60,
  2010.
\newblock Special Issue: Social Networks and Web 2.0.

\bibitem[JSK74]{jacoby1974}
Jacob Jacoby, Donald~E. Speller, and Carol~A. Kohn.
\newblock Brand choice behavior as a function of information load.
\newblock {\em Journal of Marketing Research}, 11(1):pp. 63--69, 1974.

\bibitem[KLPM10]{Kwak:2010:TSN:1772690.1772751}
Haewoon Kwak, Changhyun Lee, Hosung Park, and Sue Moon.
\newblock What is twitter, a social network or a news media?
\newblock In {\em Proceedings of the 19th international conference on World
  wide web}, WWW '10, pages 591--600, New York, NY, USA, 2010. ACM.

\bibitem[NBL10]{Naaman:2010:RMM:1718918.1718953}
Mor Naaman, Jeffrey Boase, and Chih-Hui Lai.
\newblock Is it really about me?: message content in social awareness streams.
\newblock In {\em Proceedings of the 2010 ACM conference on Computer supported
  cooperative work}, CSCW '10, pages 189--192, New York, NY, USA, 2010. ACM.

\bibitem[PG90]{Pankaj1990335}
Pankaj and Ghemawat.
\newblock The snowball effect.
\newblock {\em International Journal of Industrial Organization}, 8(3):335 --
  351, 1990.

\bibitem[PRD11]{Pollet-et-al-2011}
Thomas~V. Pollet, Sam~G.B. Roberts, and Robin~I.M. Dunbar.
\newblock Use of social network sites and instant messaging does not lead to
  increased offline social network size, or to emotionally closer relationships
  with offline network members.
\newblock {\em Cyberpsychology, Behaviour, and Social Networking},
  14(4):253--258, 2011.

\bibitem[RDL10]{ramage10microblogs}
Daniel Ramage, Susan Dumais, and Dan Liebling.
\newblock Characterizing microblogs with topic models.
\newblock In {\em Proceedings of the Fourth International Conference on Weblogs
  and Social Media (ICWSM)}, 2010.

\bibitem[RR10]{Ravikant2010}
N.~Ravikant and A.~Rifkin.
\newblock {Why Twitter Is Massively Undervalued Compared To Facebook}.
\newblock {\em TechCrunch}, 2010.
\newblock {\small
  {http://techcrunch.com/2010/10/16/why-twitter-is-massively-undervalued-compared-to-facebook/}}.

\bibitem[SG09]{Skeels:2009:SNC:1531674.1531689}
Meredith~M. Skeels and Jonathan Grudin.
\newblock {When social networks cross boundaries: A case study of workplace use
  of Facebook and LinkedIn}.
\newblock In {\em Proceedings of the ACM 2009 international conference on
  Supporting group work}, GROUP '09, pages 95--104, New York, NY, USA, 2009.
  ACM.

\bibitem[SOM10]{Sakaki:2010:EST:1772690.1772777}
Takeshi Sakaki, Makoto Okazaki, and Yutaka Matsuo.
\newblock Earthquake shakes twitter users: real-time event detection by social
  sensors.
\newblock In {\em Proceedings of the 19th international conference on World
  wide web}, WWW '10, pages 851--860, New York, NY, USA, 2010. ACM.

\bibitem[Tea11]{Facebook2011}
The Facebook~Data Team.
\newblock {Anatomy of Facebook}.
\newblock {\em Facebook}, 2011.
\newblock {\small
  {https://www.facebook.com/notes/facebook-data-team/anatomy-of-facebook/10150388519243859}}.

\bibitem[Tof84]{Toffler1984}
Alvin Toffler.
\newblock {\em {Future Shock}}.
\newblock Random House Publishing Group, 1984.

\bibitem[Tra11]{Travis2011}
Alan Travis.
\newblock {England riots: will harsher sentences act as a deterrent?}
\newblock {\em The Guardian}, 2011.
\newblock {\small
  {http://www.guardian.co.uk/uk/2011/aug/17/england-riots-harsher-sentences-deterrent}}.

\bibitem[ZG09]{Zheleva:2009:JJI:1526709.1526781}
Elena Zheleva and Lise Getoor.
\newblock To join or not to join: the illusion of privacy in social networks
  with mixed public and private user profiles.
\newblock In {\em Proceedings of the 18th international conference on World
  wide web}, WWW '09, pages 531--540, New York, NY, USA, 2009. ACM.

\bibitem[ZR09]{Zhao:2009:WPT:1531674.1531710}
Dejin Zhao and Mary~Beth Rosson.
\newblock How and why people {T}witter: the role that micro-blogging plays in
  informal communication at work.
\newblock In {\em Proceedings of the ACM 2009 international conference on
  Supporting group work}, GROUP '09, pages 243--252, New York, NY, USA, 2009.
  ACM.

\end{thebibliography}

\end{document}